\documentclass[aps,pra,superscriptaddress,twocolumn]{revtex4}

\usepackage{mathrsfs}
\usepackage{amsfonts}
\usepackage{amssymb}
\usepackage{amsmath}
\usepackage{graphicx}
\usepackage[usenames,dvipsnames]{color}
\usepackage[colorlinks=true,citecolor=blue,linkcolor=magenta]{hyperref}
\usepackage{ulem}
\usepackage{lmodern}

\newcommand{\ket}[1]{\vert{ #1 }\rangle}

\begin{document}

\title{A magic state's fidelity can be superior to the operations that created it}

\author{Ying Li}

\affiliation{Department of Materials, University of Oxford, Parks Road, Oxford OX1 3PH, United Kingdom}

\date{\today}

\begin{abstract}
The leading approach to fault tolerant quantum computing requires a continual supply of {\it magic states}. When a new magic state is first encoded, its initial fidelity will be too poor for use in the computation. This necessitates a resource-intensive {\it distillation} process that occupies the majority of the computer's hardware; creating magic states with a high initial fidelity minimises this cost and is therefore crucial for practical quantum computing. Here we present the surprising and encouraging result that raw magic states can have a fidelity significantly better than that of the two-qubit gate operations used to construct them. Our protocol exploits post-selection without significantly slowing the rate of generation and tolerates finite error rates in initialisations, measurements and single-qubit gates. This approach may dramatically reduce the size of the hardware needed for a given quantum computing task.
\end{abstract}

\pacs{}
\maketitle

\section{Introduction}

Fault tolerant quantum computing involves encoding one or more logical qubit(s) into a plurality of physical qubits and performing measurements on those physical qubits to detect and control error rates. A profound drawback of all such encodings is it is impossible to unitarily implement a universal set of operations on the logical qubits (i.e. gates) without risking the amplification of existing errors~\cite{Meier2012}. However we must of course achieve a universal set of operations in order to perform general quantum computing.

This problem can be circumvented a number of ways to perform quantum computation fault-tolerantly. Universality can be achieved by allowing a limited amplification of noise~\cite{Connor2014} or introducing additional redundancy into the code~\cite{Paetznick2013,Anderson2014,Bombin2013}. These approaches make considerable sacrifices and are not expected to tolerate as much noise as high-threshold codes, for instance the surface code~\cite{Kitaev2003,Dennis2002,Raussendorf2007PRL,Raussendorf2007NJP,Fowler2009}. Alternatively one can exploit the fact that, while the high-threshold codes do not support a complete set of fault-tolerant operations directly on our logical qubits, we can perform a more limited set of operations.

For example, in the surface code we can perform a CNOT gate between two encoded logical qubits simply by performing CNOTs between each physical qubit in one logical qubit and the corresponding physical qubit in another logical qubit. Such a procedure is called transversal. We can also perform certain other gates transversally, but crucially there are operations which we cannot achieve in this way, for example the $\pi/8$ gate. While these allowed operations do give us the ability to perform limited computations, unfortunately they do not take us beyond the algorithms that can efficiently performed on a classical computer. Therefore we need some means of upgrading the limited set of computations to a universal set, while retaining fault tolerance.

The solution for achieving universality with the surface code is the use of magic states. Suppose that we have a logical qubit $L$ encoded in a surface code composed of hundreds of physical qubits. We wish to perform a $\pi/8$ gate on $L$ in a way that is fault tolerant. Now suppose that we are {\it given} a second ancillary logical qubit $A$, this time in the magic state $\ket{0}+ e^{i\pi/4}\ket{1}$. If we now perform a CNOT controlled by $A$ targeted on $L$, and then measure out $L$ in the computational basis (which we can do transversally), then the input state on $L$ will be transferred to $A$ with the $\pi/8$ gate applied~\cite{FN1}. Given a free supply of magic states, we could consume them as-needed and thus upgrade our machine to perform full universal computing.

The question then becomes, how can we create a supply of magic states for our computer given that they are precisely the states which we {\it cannot} reach by fault tolerant operations on simple states (like logical zero). An answer is to go ahead and create `raw' magic states as well as we can, recognising that they will contain errors at an unacceptable level, and then {\it distil} those states until they are acceptable~\cite{Bravyi2005}. Distillation involves taking a large number of raw magic states and deriving a smaller number of improved states, and then repeating this as necessary until the target fidelity is reached. Crucially, this  process can be performed using only the limited set of allowed fault tolerant operations. However the process is costly in resources and the cost depends on the fidelity of the initial magic states. Consequently the distillation process may occupy the majority of the machine's hardware (in Ref.~\cite{Fowler2012} it is estimated that implementing Shor's algorithm would require a machine with over $90\%$ of qubits dedicated to magic state distillation). To minimise the hardware cost, one could think of either designing more efficient distillation algorithms~\cite{Meier2012,Bravyi2012,Jones2013,Fowler2013,Campbell2012,Campbell2014} or simply improving the initial fidelity of encoded magic states.

In this paper, we describe a highly efficient protocol for creating `raw' magic states in the surface code. As with previous authors, we take a single physical qubit in the desired magic state, and then perform a procedure that yields the same state in an encoded form. There are many such protocols for encoding a state into various topological codes. The basic idea is to grow the magic state from the physical-qubit level to the full-size-encoding level by increasing the code distance~\cite{Dennis2002,Fowler2009,Horsman2012,Raussendorf2006}. Similar ideas can be used to encode entanglement into quantum networks~\cite{Perseguers2010,Grudka2012,Li2012} or unknown states into various topological codes~\cite{Lodyga2014}.

The aim of our new protocol is to minimise the noise in the `raw' encoded magic state before any distillation is performed. Such noise is potentially induced by any imperfect operation in the encoding circuit. We begin by considering the case that the error rate for single-qubit operations is far lower than the two-qubit errors. We note that this is indeed the case in many real implementations; even in the system with the highest fidelity ever reported for a two-qubit gate, i.e.~$99.9\%$ between two trapped ions~\cite{Ballance2014}, the fidelity of single ion operations has been reported~\cite{Harty2014} at far higher levels, reaching $99.9999\%$. In our new protocol, under this practical condition we find that {\bf the infidelity in the encoded magic state is less than half the infidelity of even a single CNOT gate}. This is despite the fact that a large number of such gates are involved in the creation of the magic state. More specifically: When single-qubit operations are perfect and two-qubit gate noise is depolarizing, the rate of logical errors on the encoded magic state $p_\text{L} \sim (0.4) p_2$, where $p_2$ is the error rate of CNOT gates, i.e.~two-qubit gates. This observation is verified by numerical simulations. Presently we will also consider the effect of single-qubit noise, and we find that the logical error rate is still below the two-qubit error rate after switching on weak single-qubit noise. We use post-selection to suppress logical errors, but importantly the cost of doing so is modest: the success rate $\sim 60\%$ when $p_2 = 0.1\%$ and single-qubit operations are perfect (and it becomes more deterministic as error rates fall).

\section{Protocol}

\begin{figure}[tbp]
\centering
\includegraphics[width=0.9\linewidth]{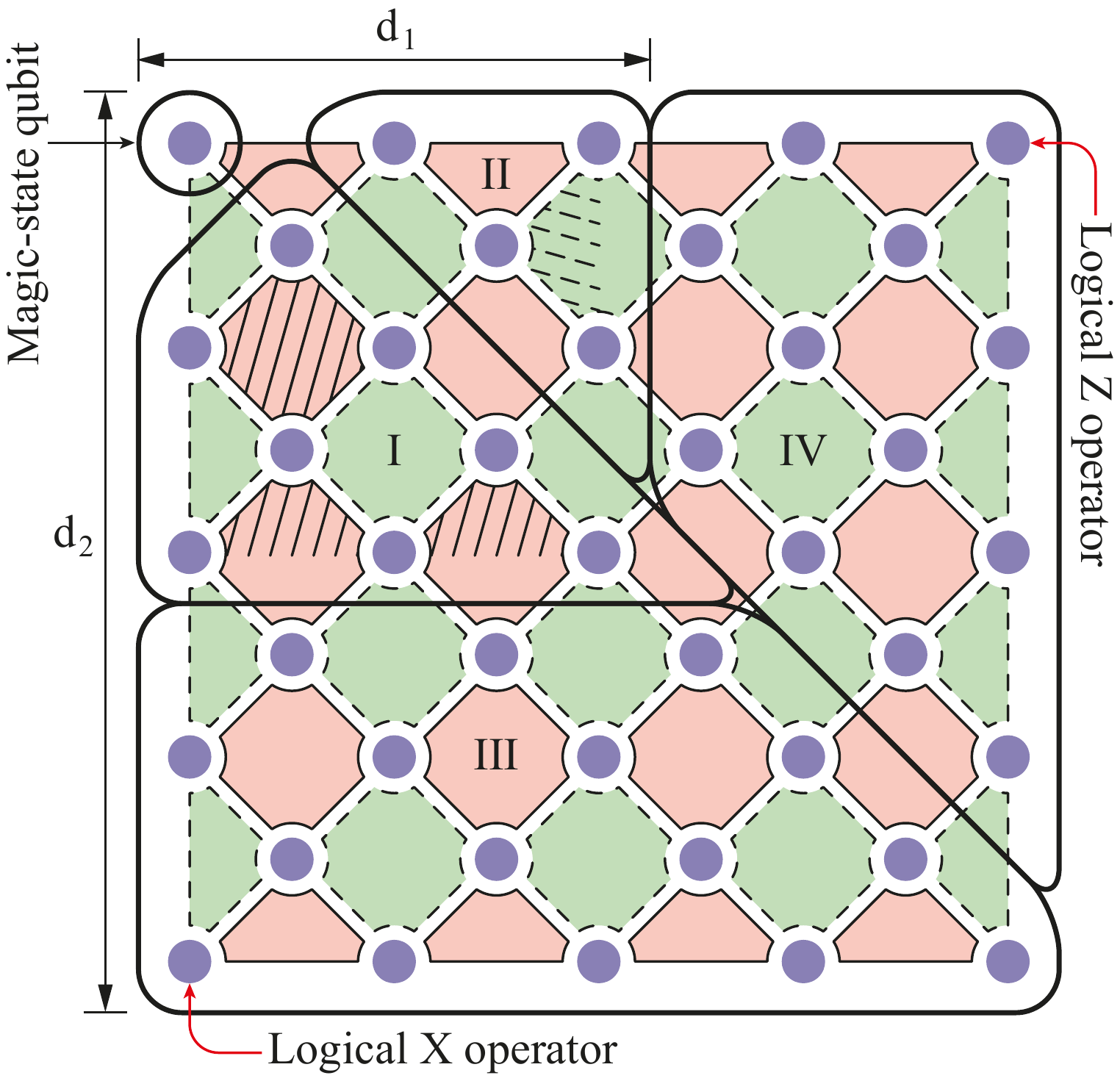}
\caption{
Schematic of the protocol for encoding a magic state into the surface code with high fidelity. Circles represent data qubits; red squares (or triangles) with solid perimeter represent $X$ stabilisers; and green squares (or triangles) with dashed perimeter represent $Z$ stabilisers. See the main text for details.
}
\label{scheme}
\end{figure}

The protocol has two phases. In the first phase, the magic state initialised on a single physical qubit is encoded into the surface code with distance $d_1$. In this stage, post selection is used to reduce the logical error rate. In the second phase, the code distance is enlarged from $d_1$ to $d_2$, the target code distance, to complete the encoding. From then on, the logical qubit is protected by correcting errors with normal syndrome detection and pairing algorithms of surface code \cite{Fowler2013}. The detailed protocol reads:

\begin{figure*}[tbp]
\centering
\includegraphics[width=0.9\linewidth]{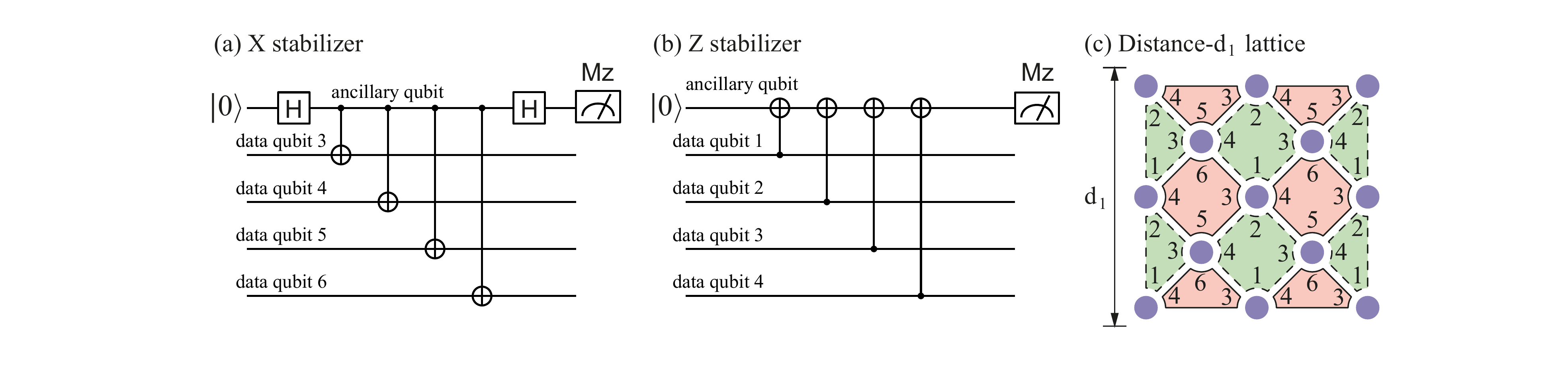}
\caption{
Circuits of stabiliser measurements in the first phase. $X$ and $Z$ stabilisers are measured with circuits in (a) and (b), respectively. The ancillary qubit is always initialised in the state $\ket{0}$ and measured in the computational basis. Optimal sequence for the CNOT gates is shown in (c). Each full round of stabiliser measurements is completed in a total of $9$ steps.
}
\label{circuit}
\end{figure*}

\begin{itemize}
\item The whole lattice of the distance-$d_2$ surface code is divided into five sets (Fig.~\ref{scheme}): the top-left corner itself, two triangular areas (I and II), and two areas (III and IV) with trapezoidal shapes. The top-left corner, area-I, and area-II form the lattice of the distance-$d_1$ surface code.
\item \textbf{First phase}:
\item[1.] A magic state is initialised on the data qubit at the top-left corner of the lattice (magic-state qubit); data qubits in area-I (area-II) are initialised in the state $\ket{+}$ ($\ket{0}$); data qubits in area-III and area-IV are not included in the first phase.
\item[2.] Stabiliser measurements are performed on the lattice of the distance-$d_1$ surface code for \textit{two} full rounds, each involving both $X$ and $Z$ stabilizer measurements. Circuits of stabiliser measurements in the first phase are shown in Fig.~\ref{circuit}, where the order of CNOT gates is designed to minimise logical errors (selecting the correct order proves to be vital to achieving a high fidelity result).
\item[3.] Error syndromes (as we define later) are detected from outcomes of stabiliser measurements. In the event that an error syndrome is found, the magic state is discarded, and all data qubits are reinitialised according to Step-1.
\item \textbf{Second phase}:
\item[4.] If no error syndrome is detected in the first phase, data qubits in area-III (area-IV) are initialised in the state $\ket{+}$ ($\ket{0}$).
\item[5.] Stabiliser measurements are performed on the entire lattice of the distance-$d_2$ surface code with any valid circuits (i.e. CNOT gates can be arranged in any convenient order that leads to valid stabilizer measurements) for one full round to complete the encoding. Regardless of whether error syndromes are found, the encoded magic state proceeds for further error correction, employing pairing algorithms and state distillation etc.
\end{itemize}

Except the magic-state qubit, all other qubits on the left (top) side of the lattice (Fig.~\ref{scheme}) are initialised in the state $\ket{+}$ ($\ket{0}$). Hence, the logical qubit is an eigenstate of $\alpha X_\text{L} + \beta Y_\text{L} + \gamma Z_\text{L}$ with the eigenvalue $+1$ if the magic-state qubit is initialised as an eigenstate of $\alpha X + \beta Y + \gamma Z$ with the eigenvalue $+1$. Here, $X_\text{L} = \prod_{i\in \text{left side}}X_i$, $Z_\text{L} = \prod_{i\in \text{top side}}Z_i$, and $Y_\text{L} = iX_{d_1}Z_{d_1}$ are Pauli operators of the logical qubit, which commute with stabiliser measurements and are conserved quantities. Therefore, the logical qubit is now in the magic state which was previously represented by the lone physical qubit in the top left.

An error syndrome is an event indicating errors. Without error, outcomes of stabiliser measurements coincide with the initialisation pattern: in the first phase, values of $X$ stabilisers in area-I and $Z$ stabilisers in area-II (stabilisers with slash lines in Fig.~\ref{scheme}) are all $+1$; similarly, in the second phase, values of $X$ stabilisers in area-I and area-III and $Z$ stabilisers in area-II and area-IV are also $+1$. Without error, the outcome of a stabiliser in later measurements is always the same as it is in the first-round measurement. Therefore, in the first phase two types of events are recognised as error syndromes: mismatches i) between the initialisation pattern and the first round of stabiliser measurements and ii) between the first and second rounds of stabiliser measurements. Error syndromes in the second phase and following stabiliser measurements are similar.

\begin{figure}[!b]
\centering
\includegraphics[width=0.9\linewidth]{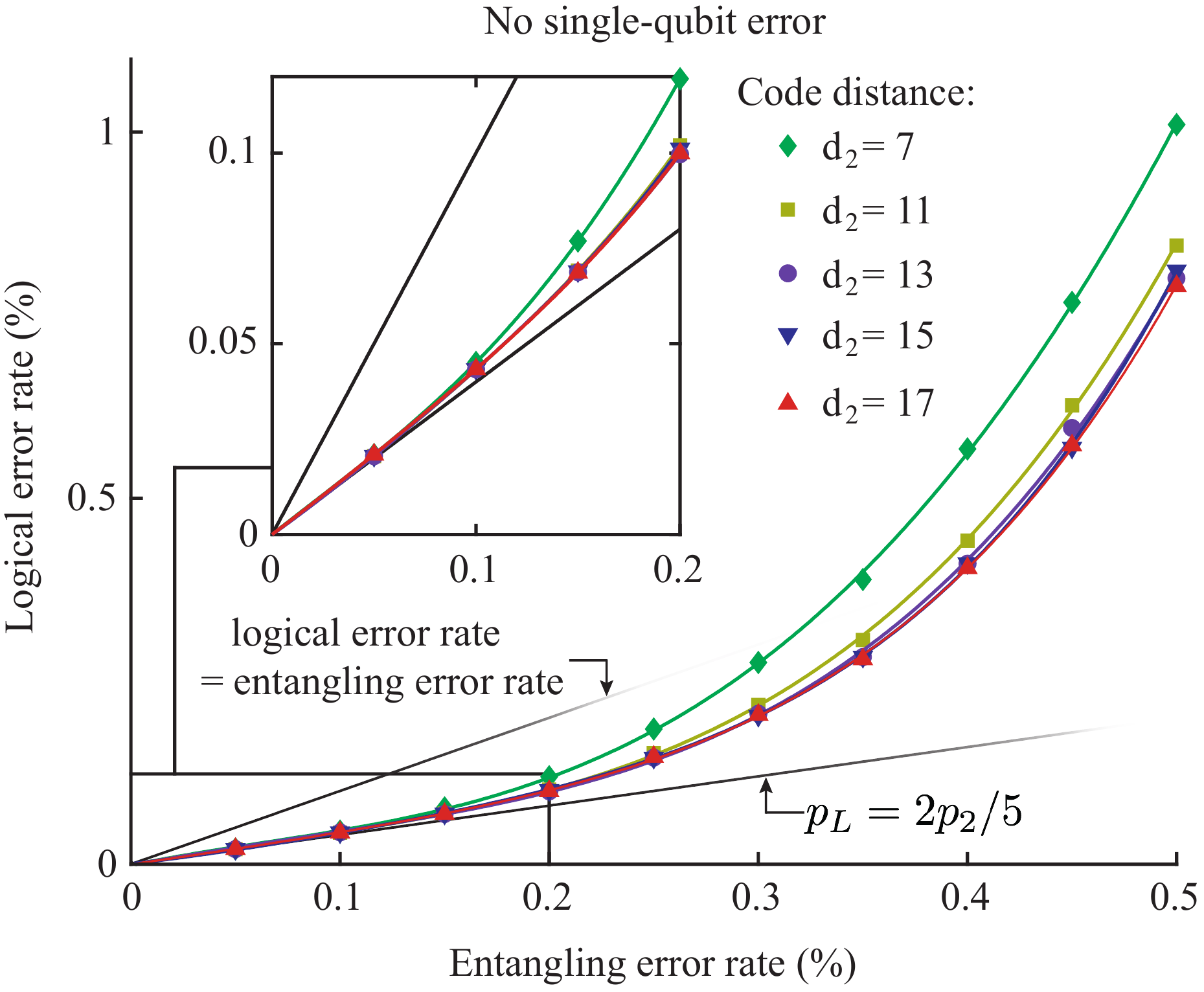}
\caption{
Error rate of the logical qubit encoded in the surface code with distance $d_2$. We have taken $d_1 = 7$ as an example. The logical error rate $p_\text{L}$ converges to the analytical limit $2p_2 /5$ when the two-qubit error rate $p_2 \lesssim 0.1\%$. This result is obtained by assuming single-qubit operations are perfect, i.e.~$p_\text{I},p_M,p_1 = 0$. Logical errors are detected after performing $d_2$ rounds of full-size stabiliser measurements (including the one in the second phase) and correcting errors with Edmonds's minimum weight matching algorithm~\cite{Kolmogorov2009}, so that short error chains are sufficiently considered in our simulations. In the second phase, we have used stabiliser-measurement circuits proposed in Ref.~\cite{Fowler2009}.
}
\label{no1qerror}
\end{figure}

Optimal circuits of stabiliser measurements (Fig.~\ref{circuit}) in the first phase are obtained by minimising logical errors on the encoded magic state. Generally, a stabiliser measurement includes an ancillary qubit and four (or three) CNOT gates between the ancillary qubit and relevant data qubits. Classifying CNOT gates by stabilisers and their orientations, there are eight sets of CNOT gates in each full round of stabiliser measurements. After searching in all valid stabiliser-measurement circuits restricted to those implementing each set of CNOT gates in parallel, we find that the logical error rate ranges from $\sim 2p_2 /5$ to $\sim 4p_2 /3$ depending on ordering. The circuit shown in Fig.~\ref{circuit} is one of the circuits providing the minimised logical errors. Given this optimal circuit we then allow for finite error rates in other operations besides the two-qubit CNOT gates; the consequences are described in the following section.

\section{Results}

\begin{figure}[tbp]
\centering
\includegraphics[width=0.9\linewidth]{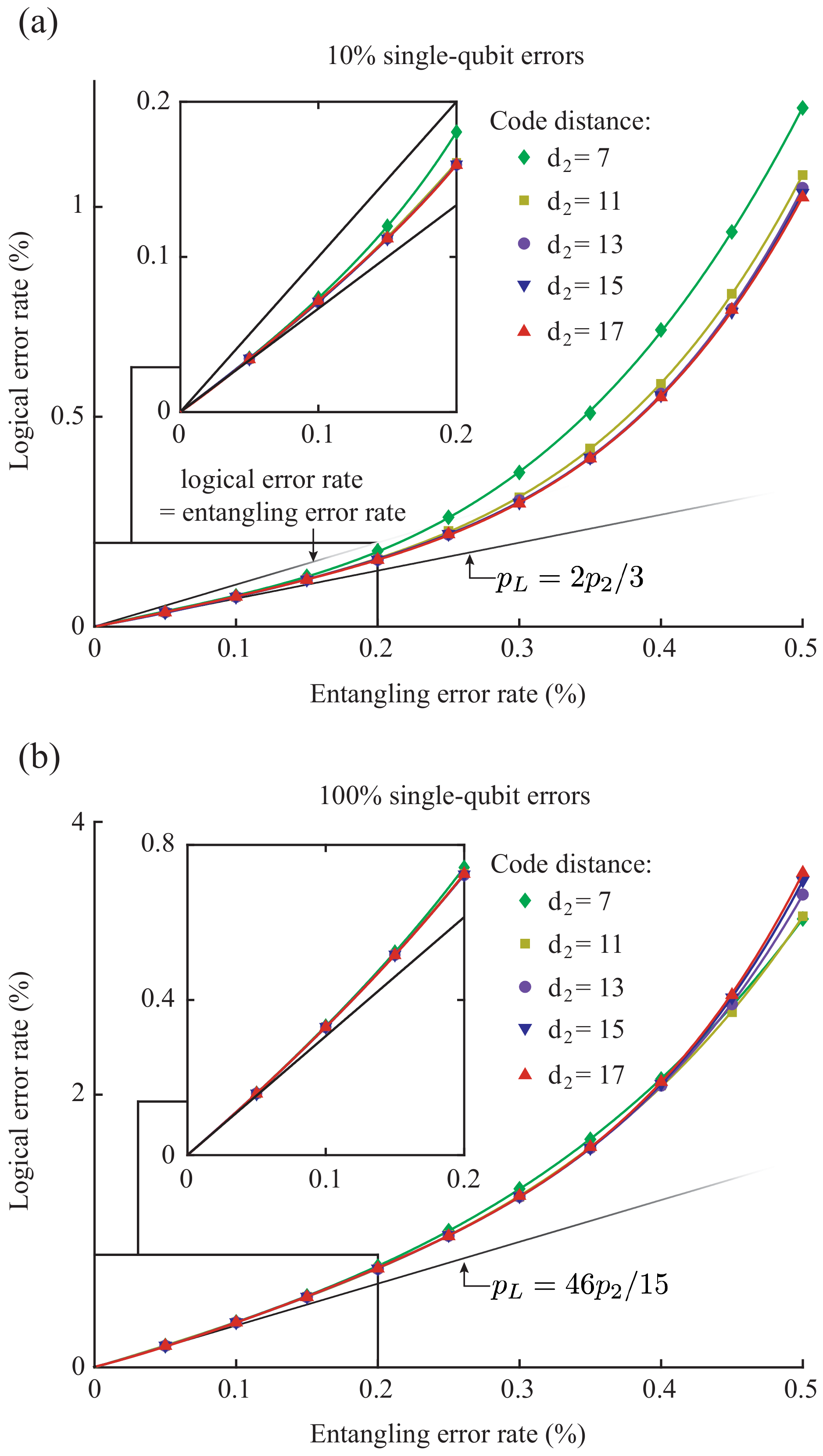}
\caption{
Logical error rate with (a) single-qubit errors $p_\text{I},p_\text{M},p_1 = p_2/10$ and (b) single-qubit errors $p_\text{I},p_\text{M},p_1 = p_2$. With these $10\%$ ($100\%$) single-qubit errors, the logical error rate converges to $2p_2 /3$ ($46p_2 /15$) consistent with Eq.~(\ref{eq:pL}).
}
\label{with1qerrors}
\end{figure}

Operations on physical qubits which are included in our protocol are: initialisation in the state $\ket{0}$; measurement in the computational basis ($\ket{0}$ and $\ket{1}$); single-qubit gates; and CNOT gate. We assume a qubit may be initialised in the incorrect state $\ket{1}$ with the probability $p_\text{I}$; the measurement may report an incorrect outcome with the probability $p_\text{M}$; and each single-qubit gate and CNOT gate (i.e.~two-qubit gate) may induce an error with the probability $p_1$ and $p_2$ respectively. A noisy gate is modelled as a perfect gate followed by single-qubit depolarizing noise for single-qubit gates and two-qubit depolarizing noise for the CNOT gate~\cite{Raussendorf2007PRL}.

The logical qubit is sensitive to noise when the code distance is small and more stable when the code distance is larger. Therefore, most of logical errors occur in the first phase. Utilizing post selection and optimised stabilizer-measurement circuits, errors in the first phase are well suppressed. With the depolarizing error model, the rate of logical errors on the encoded magic state is:
\begin{eqnarray}
p_\text{L} = \frac{2}{5}p_2 + 2p_\text{I} + \frac{2}{3}p_1 + \mathcal{O}(p^2).
\label{eq:pL}
\end{eqnarray}
Initialisation errors on the magic-state qubit and the next data qubit on the same horizontal line (second data qubit) can cause logical errors occurring with the probability $2p_\text{I}$. The single-qubit gate for rotating the magic-state qubit to the magic state may induce a logical error with the probability $2p_1/3$. CNOT gates in the first round of stabiliser measurements may also induce logical errors. There are $6$ kinds of CNOT-gate errors that can result in logical errors, and each of them occurs with the probability $p_2/15$. These errors are $[Z_c]$, $[X_cX_t]$ and $[Y_cX_t]$ induced by the CNOT gate on the magic-state qubit for measuring the $Z$ stabiliser, $[Z_cZ_t]$ induced by the CNOT gate on the magic-state qubit for measuring the $X$ stabiliser, and $[X_cX_t]$ and $[Y_cX_t]$ induced by the CNOT gate on the second data qubit for measuring the $Z$ stabiliser. Here, $c$ and $t$ respectively denote the control and target qubits in corresponding CNOT gates. All other errors do not cause logical errors solely hence contribute to the logical error rate in second order.

This analytical result is verified by numerical simulations, example curves are shown in Fig.~(\ref{with1qerrors}) where we take $d_1 = 7$ as an example. In general, a larger $d_1$ implies a higher fidelity but also a smaller success probability. By choosing $d_1=7$, we find that logical errors are well suppressed and at the same time the success probability is still high. When single-qubit operations are perfect ($p_\text{I},p_\text{M},p_1 = 0$), the logical error rate converges to $2p_2 /5$ as shown in Fig.~\ref{no1qerror}. After switching on all single-qubit noise to $10\%$, i.e.~$p_\text{I},p_\text{M},p_1 = p_2/10$, the logical error rate increases to $2p_2 /3$ as shown in Fig.~\ref{with1qerrors}(a). Thus the logical error rate remains lower than the physical two-qubit error rate even when other error sources are present at a finite level (and indeed in many physical implementations there is more than an order of magnitude separating the two-qubit and single-qubit error rates). Ultimately however if all forms of single-qubit operation suffer error rates equal to the two-qubit error rate, then logical error rate does exceed this common physical error rate and reaches $46p_2 /15$ [see Fig.~\ref{with1qerrors}(b)].

With $d_1 = 7$ and two-qubit error rate $p_2 = 0.1\%$, the first phase succeeds with a probability in the range $0.38 \sim 0.59$ depending on the rate of single-qubit errors. However, by adaptive use of hardware resources the protocol's effective success rate is much higher: For practical quantum computation, the target surface code usually has a large distance $\sim 20$. If we choose $d_2 = 21$, the entire lattice can be divided into $9$ copies of $d_1$ lattice, hence the first phase can be attempted in parallel, and the rate of obtaining at least one success is $\sim 1-(0.5)^9 \simeq 0.998$. Although the successful copy may not be the one located at the top-left corner, we still can enlarge the code distance from $d_1$ to $d_2$ by adapting the initialisation pattern in the second phase. Because Eq.~(\ref{eq:pL}) is only determined by the first phase, the overall fidelity will not be affected significantly.

Finally as an aside we note that the protocol described here can also be used to encode magic states to a punctured surface code~\cite{Fowler2009}.

\section{Conclusions}

We have proposed a new protocol for encoding magic states into the surface code with high-fidelity. Remarkably, we find that the optimal gate sequence results in noise on the encoded magic state which is {\it lower than half of the noise induced by a single physical CNOT gate}. Compared with the previous protocol~\cite{Raussendorf2006}, logical errors due to two-qubit noise are reduced by about a factor of ten. This can profoundly reduce the size of the hardware needed for quantum computing: For example with the 15-to-1 distillation protocol~\cite{Bravyi2005} the logical error rate can be reduced from $p_\text{in}$ for input magic states to $p_\text{out} = 35p_\text{in}^3$ (for small $p_\text{in}$) for the output magic state for each round of distillation, i.e.~the advantage of our protocol is then a factor of $10^{3n}$ after $n$ rounds of distillations. We can expect that this will reduce the required number of rounds by one (as, for example, if $p_2 = 0.1\%$ and the target error rate of the distillation is anywhere between $10^{-15}$ and $10^{-24}$). The hardware requirements can then be reduced by a factor of $15$. Given the anticipated expense and complexity of quantum computing devices, we believe this is an important and very encouraging result.

\begin{acknowledgments}
I wish to thank Simon Benjamin, Earl Campbell, Austin Fowler, Clare Horsman, and Naomi Nickerson for helpful discussions. I am also grateful to Simon Benjamin and Earl Campbell for their help in preparing the introductory parts of this manuscript.
\end{acknowledgments}


\begin{thebibliography}{9}
\bibitem{FN1} In fact one of two different gates on the input state will have occurred, depending on the outcome of the measurement -- but these gates are related by a rotation that can be performed fault-tolerantly.
\end{thebibliography}
\end{document}